\documentstyle[amssymb,aps]{revtex}

\begin{document}
\author{Sviatoslav S. Sokolov$^{1,2}$ and Nelson Studart$^1$}
\address{$^{1}$Universidade Federal de S\~{a}o Carlos, 13565-905 S\~{a}o Carlos, SP,\\
Brazil\\
$^2$B. I. Verkin Institute for Low Temperature Physics and Engineering,\\
National Academy of Sciences of Ukraine, 310164 Kharkov, Ukraine}
\title{Coupled phonon-ripplon modes in a single wire of electrons on the
liquid-helium surface}
\maketitle
\date{}

\begin{abstract}
The coupled phonon-ripplon modes of the quasi-one-dimensional electron chain
on the liquid helium sutface are studied. It is shown that the
electron-ripplon coupling leads to the splitting of the collective modes of
the wire with the appearance of low-frequency modes and high-frequency
optical modes starting from threshold frequencies. The effective masses of
an electron plus the associated dimple for low frequency modes are estimated
and the values of the threshold frequencies are calculated. The results
obtained can be used in experimental attempts to observe the phase
transition of the electron wire into a quasi-ordered phase.

PACS: 73.20.Dx; 73.90.+f ; 63.20.Dj\vspace{0.6cm}
\end{abstract}

One of the most distinctive features of the quasi-two-dimensional (Q2D)
classical electron system localized on liquid helium surface is that it
exhibits liquid-solid like phase transition at low enough temperature $T$
and high electron densities $n_{s}$ when the interaction coupling $\Gamma
=e^{2}\sqrt{\pi n_{s}}/T\simeq 140$. The peculiarities of the transition
leading to the formation of 2D Wigner solid (WS) are widely discussed both
theoretically and experimentally for Q2D surface electrons.\cite{general}
Due to their interaction with liquid surface oscillations (ripplons),
electrons with 2D wave vector ${\bf k}$ are coupled to ripplons with ${\bf q}%
={\bf G}+{\bf k}$, where ${\bf G}$ belongs to the 2D reciprocal lattice, and
crystallize in a triangular lattice whose elementary cell contains the
electron and the self-induced surface deformation (dimple). As a
consequence, the spectrum of collective modes of the WS on the helium
surface differs significantly from the collective mode spectrum of an ideal
2DWS of electrons decoupled from ripplons\cite{bm77} and splits into
low-frequency and high-frequency modes.\cite{general,fhp79,ms83}
Low-frequency longitudinal and transverse modes are acoustic-like and are
related to in-phase motion of the electron and associated dimple.
High-frequency optical modes start from $\omega _{\alpha }$ at ${\bf k}=0$
and correspond to the relative oscillations of electrons and dimples. The
threshold values $\omega _{\alpha }$ are determined by the strength of the
electron-ripplon interaction and are significantly larger than typical
frequencies of ripplons involved in electron-ripplon scattering processes.
This allows us to average the Hamiltonian of the electron-ripplon
interaction only over high-frequency modes. Note that the observation of
resonances in the absorption spectrum at peculiar coupled phonon-ripplon
mode frequencies leads, for the first time, to the evidence of the electron
WS over the liquid helium surface.\cite{ga79}

Last years there is an increasing interest in investigating
quasi-one-dimensional (Q1D) electron systems over liquid helium.\cite
{geom,eletr,revss97} In such systems the electrons are confined not only in
the direction normal to the liquid surface (where the 2D electron system
takes place) but also by a lateral constriction, which is provided by
geometric or electrostatic means, leading to the creation of a classical
wire system (low-density regime) similar to quantum wires in semiconductor
heterostructures. The confinement potential can be, within a good accuracy,
approximated by a parabolic potential. Figure 1 illustrates an example of
the realization of the single electron wire by using a bent foil that
provides the helium surface with a curvature radius $R\sim 10^{-3}-10^{-4}$
cm.\cite{q1dexp} If the holding electric field $E_{\perp }$ along the $z$
axis is $\sim 10^{3}$ V/cm, the electron wave function for the motion across
the channel becomes oscillatory-like with frequency $\omega _{0}^{2}=$ $%
eE_{\perp }/mR$. It is essential to emphasize that, despite the additional
restriction in the electron motion, the electron-ripplon interaction can
still be considered in the same manner as in 2D case.\cite{shs95}

The possibility that a linear chain of electrons, with period $a$ , may
undergo a phase transition into a ordered state is not obvious as in 2D. It
is well known that there is no true long order in a strictly 1D infinite
system. However, one can expect long enough regions of a quasi-ordered
electron chain in these Q1D electrons for very low temperatures.\cite
{chaplik} Based on this assumption, the spectrum of longitudinal and
transverse phonons was obtained in the case of the decoupled electron chain
from the solution of equations of motion in the harmonic approximation.\cite
{chaplik,slava} The dispersion laws for longitudinal ($\omega _{l})$ and
transverse ($\omega _{t})$ branches are given by\cite{slava} 
\begin{equation}
\omega _{l}^{2}=\omega _{e}^{2};\ \ \text{and}\ \ \omega _{t}^{2}=\omega
_{0}^{2}-\frac{\omega _{e}^{2}}{2}  \label{1}
\end{equation}
where the frequency $\omega _{e}$, for $|k_{x}a|\ll 1$, can be written as 
\[
\omega _{e}^{2}=\frac{4e^{2}}{ma}\sum_{n\geq 1}\left( \frac{1-\cos nak_{x}}{%
n^{3}}\right) \simeq \frac{2e^{2}}{ma}k_{x}^{2}\ln \left( \frac{1}{|k_{x}a|}%
\right) . 
\]
Here $k_{x}$ is the 1D wave vector along the wire. As it can be seen from
Eq. (1), the transverse branch is stable only if the lateral confinement
exists. As it was shown in Ref. \cite{slava}, the stability holds for $%
E_{\perp }>7\zeta (3)eR/2a^{3}$, where $\zeta (x)$ is the Riemann's
zeta-function. For $a=10^{-4}$ cm and $R=5\times 10^{-4}$ cm, one has $%
E_{\perp }>300$ V/cm. The spectrum of longitudinal oscillations of the
classical Q1D itinerant electron system was also studied by Sokolov and
Studart\cite{ss98} using the random-phase approximation. The results show
qualitatively similar dispersion laws as in Eq. (1) when $a$ in the argument
of the logarithm is replaced by the scale of the electron localization
across the wire. No electron-ripplon coupling was taken into accounted in
Refs. 10-12, even though, as stated previously, such a coupling should
influence, in a crucial way, the mode spectrum of the Q2DWS.\cite{ga79}

In this communication, we discuss the possibility of observation of the
transition into a quasi-ordered state of the classical Q1D electron system
over liquid helium by calculating the coupled phonon-ripplon modes. We show
that the formation of an ordered state in the single electron wire should be
accompanied by drastic changes in the mode spectrum in comparison with that
of the decoupled WS wire (Eq. (1)). New unusual branches appear which can be
checked experimentally.

The Hamiltonian of the electron wire is given by

\begin{equation}
\hat{H}=\hat{H}_{e}+\hat{H}_{r}+\hat{H}_{er},  \label{2}
\end{equation}
where the $N$-electron Hamiltonian for small displacements from the
equilibrium positions $\{x_{l}^{(0)};0;0\}$, $l$ $=1,2...N$, can be written
as

\begin{equation}
\hat{H}_{e}=\frac{m}{2}\sum_{l=1}^{N}(\stackrel{\cdot }{u}_{xl}^{2}+%
\stackrel{\cdot }{u}_{yl}^{2}+\omega _{0}^{2}u_{yl}^{2})+\frac{e^{2}}{2a^{3}}%
\sum_{l\neq j}\left[ \frac{\left( u_{xl}-u_{xj}\right) ^{2}}{%
|x_{l}^{(0)}-x_{j}^{(0)}|^{3}}-\frac{\left( u_{yl}-u_{yj}\right) ^{2}}{%
2|x_{l}^{(0)}-x_{j}^{(0)}|^{3}}\right] .  \label{3}
\end{equation}
Here $u_{xl}$ and $u_{yl}$ are the electron displacements along $x$ and $y$
directions, respectively. In terms of the Fourier-transformed phonon
displacements $\zeta _{xk}$ and $\zeta _{yk},$ the Hamiltonian can be
expressed as

\begin{equation}
\hat{H}_e=\frac m2\sum_k\left[ |\stackrel{\cdot }{\zeta }_{xk}|^2+|\stackrel{%
\cdot }{\zeta }_{yk}|^2+\omega _e^2|\zeta _{xk}|^2+\left( \omega _0^2-\frac{%
\omega _e^2}2\right) |\zeta _{yk}|^2\right] .  \label{5}
\end{equation}

The Hamiltonian for free ripplons is given in the standard form

\begin{equation}
\hat{H}_{r}=\frac{1}{2}\sum_{{\bf q}}\left( \frac{\rho }{q}\right) \left( |%
\stackrel{\cdot }{\xi }_{{\bf q}}|^{2}+\omega _{q}^{2}|\xi _{{\bf q}%
}|^{2}\right) ,  \label{6}
\end{equation}
where $\xi _{{\bf q}}$ is Fourier-transformed displacement of the liquid
surface from the equilibrium position at $z=0$, and $\omega _{q}^{2}\simeq
(\alpha /\rho )q^{3}+\stackrel{\thicksim }{g}q$. The helium bulk density and
the surface tension coefficient are denoted $\rho $ and $\alpha $
respectively, and $\stackrel{\thicksim }{g}$ is the gravity acceleration.

The Hamiltonian $\hat{H}_{er}$ describes the electron-ripplon interaction
and can be written as

\begin{equation}
\hat{H}_{er}=\frac{1}{\sqrt{S}}\sum_{l=1}^{N}\sum_{{\bf q}}\xi _{{\bf q}}V_{%
{\bf q}}\exp \left( -q^{2}\langle u_{f}^{2}\rangle /4\right)
e^{iq_{x}x_{l}^{(0)}}e^{(q_{x}u_{xl}+q_{y}u_{yl})},  \label{7}
\end{equation}
where $S$ is the surface area, $V_{{\bf q}}$ is the electron-ripplon
interaction averaged over the ground state electron wave function along the $%
z$ direction.\cite{interac} The Debye-Waller factor $\exp \left(
-q^{2}\langle u_{f}^{2}\rangle /4\right) ,$ which appears in Eq. (\ref{7}),
comes from taking into account the ``smearing'' of the electron positions
due to high-frequency modes and corresponds to the contribution $\langle
u_{f}^{2}\rangle $ in the electron root-mean-square (RMS) displacements from
the fast modes. So, $u_{xl}$ and $u_{yl}$ in Eq. (\ref{7}) should be
considered as the electron displacements at low enough velocities. Expanding
the exponential function in Eq. (\ref{7}) up to quadratic terms and making
the Fourier transform of the electron displacements one can finally arrive
to the following expression for the total Hamiltonian:

\begin{eqnarray}
H &=&\sum_{k}\{\frac{1}{2}\sum_{{\bf q}g}\left( \frac{\rho }{q}\right)
\left( |\stackrel{\cdot }{\xi }_{{\bf q}}|^{2}+\omega _{q}^{2}|\xi _{{\bf q}%
}|^{2}\right) \delta _{q_{x};g+k_{x}}  \label{8} \\
&&+\frac{m}{2}\left[ |\stackrel{\cdot }{\zeta }_{xk_{x}}|^{2}+|\stackrel{%
\cdot }{\zeta }_{yk_{x}}|^{2}+\left( \omega _{e}^{2}+\omega _{dx}^{2}\right)
|\zeta _{xk_{x}}|^{2}+\left( \omega _{0}^{2}+\omega _{dy}^{2}-\frac{\omega
_{e}^{2}}{2}\right) |\zeta _{yk_{x}}|^{2}\right]  \nonumber \\
&&+i\sqrt{n_{s}}\sum_{{\bf q}g}V_{{\bf q}}\exp \left( -q^{2}\langle
u_{f}^{2}\rangle /4\right) \left( q_{x}\zeta _{x,-k_{x}}+q_{y}\zeta
_{y,-k_{x}}\right) \xi _{{\bf q}}\delta _{q_{x};g+k_{x}}\}.  \nonumber
\end{eqnarray}
Here $\delta _{\alpha \beta }$ is the Kronecker symbol, $g=(2\pi /a)n$ is
the 1D reciprocal-chain vector with $n=0,\pm 1,\pm 2,...\pm N/2$. As it is
seen in Eq. (\ref{8}), the phonon displacements $\zeta _{x(y)k_{x}}$ are
coupled to ripplons with $q=\sqrt{(g+k_{x})^{2}+q_{y}^{2}}.$ The frequencies 
$\omega _{dx}$ and $\omega _{dy}$ are

\begin{equation}
\omega _{dx}^{2}=\sum_{gq_{y}}C_{gq_{y}}\omega _{gq_{y}}^{2}\quad \text{%
and\quad }\omega _{dy}^{2}=\sum_{gq_{y}}C_{gq_{y}}\left( \frac{q_{y}}{g}%
\right) ^{2}\omega _{gq_{y}}^{2},  \label{9}
\end{equation}
where the coefficients

\[
C_{gq_{y}}=\frac{n_{s}V_{gq_{y}}^{2}g^{2}\exp \left( -\frac{%
(g^{2}+q_{y}^{2})\langle u_{f}^{2}\rangle }{4}\right) }{\alpha
m(g^{2}+q_{y}^{2})\omega _{gq_{y}}^{2}} 
\]
are dimensionless parameters related to the strength of the electron-ripplon
coupling. The double index $gq_{y}$ in Eq. (\ref{9}) means that $g$
substitutes $q_{x}$ in the quantities depending on $q=\sqrt{%
q_{x}^{2}+q_{y}^{2}}$.

The dispersion laws of coupled phonon-ripplon oscillations can be obtained
by solving the equations of motion for the Hamiltonian given in Eq. (\ref{8}%
). In the long wave limit, $k_{x}\ll \sqrt{g^{2}+q_{y}^{2}}$, the dispersion
equations can be written as

\begin{equation}
Z_{l}\omega ^{2}-\omega _{e}^{2}=0\quad \text{and\quad }Z_{t}\omega
^{2}-\omega _{0}^{2}+\frac{\omega _{e}^{2}}{2}=0.  \label{10}
\end{equation}
where

\[
Z_{l}=1+\sum_{gq_{y}}C_{gq_{y}}\frac{\omega _{gq_{y}}^{2}}{\omega
_{gq_{y}}^{2}-\omega ^{2}}\quad \text{and\quad }Z_{t}=1+%
\sum_{gq_{y}}C_{gq_{y}}\left( \frac{q_{y}}{g}\right) ^{2}\frac{\omega
_{gq_{y}}^{2}}{\omega _{gq_{y}}^{2}-\omega ^{2}}. 
\]
One can easily see that the above equations have solutions that differ
substantially from the phonon modes given by Eq. (\ref{1}). We consider both
low and high-frequency regions of the spectrum. In the limit of $\omega
^{2}\ll \omega _{gq_{y}}^{2}$, the phonon-like solutions of Eq. (\ref{10})
can be given by 
\begin{mathletters}
\begin{equation}
\stackrel{\thicksim }{\omega }_{l}\simeq \left( \frac{m}{M_{l}}\right)
^{1/2}\omega _{l}\quad \text{and\quad }\omega _{t}\simeq \left( \frac{m}{%
M_{t}}\right) ^{1/2}\omega _{t},  \label{11a}
\end{equation}
which are formally equivalent to the modes expressed in Eq. (\ref{1}), but
depend on the effective masses of the electron plus dimple instead on the
free electron mass as it follows:

\begin{equation}
M_{l}=m\left[ 1+\sum_{gq_{y}}C_{gq_{y}}\right] \quad \text{and\quad }%
M_{t}=m\left[ 1+\sum_{gq_{y}}C_{gq_{y}}\left( \frac{q_{y}}{g}\right)
^{2}\right] .  \label{11b}
\end{equation}
Note that the longitudinal mode is gapless and the spectrum is similar to
that of low frequency modes in the coupled 2DWS over liquid helium at same $%
q $. However, in contrast to the 2D case, the dispersion of longitudinal and
transverse modes shows different effective masses. For large holding fields,
where $V_{gq_{y}}\simeq eE_{\perp }$, and summing over $q_{y}$, analytical
expressions for the effective masses are obtained as

\end{mathletters}
\begin{mathletters}
\begin{equation}
M_{l}\simeq m\left[ 1+\frac{n_{l}e^{2}E_{\perp }^{2}\rho \langle
u_{f}^{2}\rangle ^{2}}{4\sqrt{\pi }\alpha ^{2}m}\sum_{g>0}g^{2}\exp
(-g^{2}\langle u_{f}^{2}\rangle /2)\Psi \left( \frac{5}{2};3;\frac{%
g^{2}\langle u_{f}^{2}\rangle }{2}\right) \right] ,  \label{12a}
\end{equation}

\begin{equation}
M_{t}\simeq m\left\{ 1+\frac{n_{l}e^{2}E_{\perp }^{2}\rho \langle
u_{f}^{2}\rangle }{4\sqrt{\pi }\alpha ^{2}m}\left[ \frac{1}{\sqrt{\pi }}\Psi
\left( 2;2;\frac{\kappa _{c}^{2}\langle u_{f}^{2}\rangle }{2}\right) +\sum_{%
\dot{g}>0}\exp (-g^{2}\langle u_{f}^{2}\rangle /2)\Psi \left( \frac{5}{2};2;%
\frac{g^{2}\langle u_{f}^{2}\rangle }{2}\right) \right] \right\} .
\label{12b}
\end{equation}
Here $\kappa _{c}=(\rho \stackrel{\thicksim }{g}/\alpha )^{1/2}$ is the
helium capillary constant, $n_{l}=N/L_{x}$ is the linear electron density
along the wire with $L_{x}$ the system length, and $\Psi (a;b;x)$ is the
degenerated hypergeometric Tricomi's function. Equations (11) were obtained
under the condition $k_{c}\ll g_{1}$.

In order to estimate the magnitudes of the effective masses we need
characteristic values of $a$, $E_{\perp },$ and $\langle u_{f}^{2}\rangle .$
The relationship between $n_{l}$ and $E_{\perp }$ is much more complicated
than its 2D counterpart, and we consider here $a\sim n_{l}^{-1}\sim 1/\sqrt{%
n_{s}}=10^{-4}$ cm and the high-field limit of $E_{\perp }=3000$ V/cm,
because we neglected the contribution of the polarization interaction
between the electron and the liquid helium in Eqs. (11). The RMS
contribution from the fast modes $\langle u_{f}^{2}\rangle $ can be treated
as an fitting parameter as in Ref. \cite{fhp79} or can be calculated
rigorously in a self-consistent procedure.\cite{ms83} Unfortunately there is
no experimental data up to now and a self-consistent calculation of $\langle
u_{f}^{2}\rangle $ in the Q1D case is cumbersome. On the grounds of the
results of Ref.\cite{ms83}, we choose $\langle u_{f}^{2}\rangle =\hslash
/(2m\omega _{dx})\coth (\hslash \omega _{dx}/2T)$ and considering the regime
where $2T\ll \hbar \omega _{dx},$ we find best self-consistency by taking $%
\sqrt{\langle u_{f}^{2}\rangle }\simeq 10^{-1}a\ll a.$ Furthermore, this
value also satisfies well the condition $\sqrt{\langle u_{f}^{2}\rangle }\ll
R=5\times 10^{-4}$cm.

For realistic values of $a=10^{-4}$ cm and $E_{\perp }=3000$ V/cm, $%
M_{l}=2.3\times 10^{4}\;m$ which is near two orders of magnitude larger than
the effective mass in the low-frequency mode in the 2DWS for $n_{s}\gtrsim
10^{8}$.\cite{general} $M_{t}$ is significantly larger than $M_{l}$ due to
the large contribution of $g=0$. Note the $\kappa _{c}$-dependent argument
of Tricomi's function in the first term in square brackets of Eq. (\ref{11b}%
). This contribution is formally related with transverse oscillations as the
whole electron-dimple chain when $k_{x}\rightarrow 0$ and the interparticle
distance becomes very large ($g\rightarrow 0$). Such uniform oscillations
yield large $M_{t}$, and $M_{t}\rightarrow \infty $ for $k_{c}=0$. However, $%
k_{c}\neq 0$ leads to very large, even though finite, value of $M_{t}$ when
the condition $k_{c}\gg a$ plays the role of an effective cutoff in the
divergent contribution of the $g=0$ term in Eq. (\ref{12b}). On the other
hand, all electron positions are equivalent under longitudinal displacements
of the whole electron chain and, because this, the term with $g=0$ does not
contribute to the longitudinal oscillations of the chain at very small $%
k_{x} $. So, $M_{l}$ is significantly smaller than $M_{t}$. For the same
values of $a$ and $E_{\perp }$, we obtain $M_{t}\simeq 1.1\times 10^{11}\;m$
such that the threshold frequency $\stackrel{\thicksim }{\omega }%
_{0}=(m/M_{t})^{1/2}\omega _{0}$ of the transverse mode $\stackrel{\thicksim 
}{\omega }_{t}$ is strongly softened and decreases more than 5 orders of
magnitude in comparison with $\omega _{0}\simeq 10^{11}$ s$^{-1}$ for $%
E_{\perp }=3000$ V/cm and $R=5\times 10^{-4}$ cm.\cite{shs97} From these
estimates, one can conclude that the electron-ripplon coupling in the WS
wire is stronger than in the 2DWS.

In the limit of high frequencies, $\omega ^{2}\gg \omega _{gq_{y}}^{2}$, the
solutions of Eqs. (\ref{10}) are two optical branches

\end{mathletters}
\begin{equation}
\omega _{tr(l)}^{2}\simeq \omega _{dx}^{2}+\omega _{e}^{2}\quad \text{%
and\quad }\omega _{tr(t)}^{2}\simeq \omega _{dy}^{2}+\omega _{0}^{2}-\frac{%
\omega _{e}^{2}}{2},  \label{13}
\end{equation}
where the frequencies $\omega _{dx}$ and $\omega _{dy}$, given by Eq. (\ref
{9}), can be expressed, in the strong-field limit of $V_{gq_{y}}\simeq
eE_{\perp },$ as 
\begin{mathletters}
\begin{equation}
\omega _{dx}^{2}\simeq \frac{n_{l}e^{2}E_{\perp }^{2}}{\alpha m}%
\sum_{g>0}g\left[ 1-%
\mathop{\rm erf}%
\left( \sqrt{g^{2}\langle u_{f}^{2}\rangle /2}\right) \right] ,  \label{14a}
\end{equation}

\begin{equation}
\omega _{dy}^{2}\simeq \frac{n_{l}e^{2}E_{\perp }^{2}}{\alpha m\sqrt{2\pi
\langle u_{f}^{2}\rangle }}\sum_{g}\exp (-g^{2}\langle u_{f}^{2}\rangle
/2)-\omega _{dx}^{2}.  \label{14b}
\end{equation}
We remind that in the 2DWS both optical branches start from the same
threshold frequency $\omega _{\alpha }$ at $q=0$. For $a=10^{-4}$ cm and $%
E_{\perp }=3000$ V/cm, the threshold frequencies of the Q1D electron chain
are $\omega _{dx}\simeq 7.1\times 10^{9}$ s$^{-1}$ and $\omega _{dy}\simeq
7.5\times 10^{9}$ s$^{-1}.$ We observe that $\omega _{dy}$ is significantly
smaller than $\omega _{0}$ for the same $E_{\perp }$ and the threshold
frequency of $\omega _{tr(t)}$ in Eq. (\ref{13}) practically coincides with $%
\omega _{0}.$ So, this mode branch is almost unaffected by the
electron-ripplon coupling (see Eq. (\ref{1})).

Besides phonon-like modes given by Eqs. (11) and (\ref{13}), the Eq. (\ref
{10}) have also solutions which represent optical modes whose dispersion
laws starting from frequencies which are only rather lower than $\omega
_{gq_{y}}$ at $k_{x}=0.$ By increasing $k_{x}$, the frequencies of these
modes practically reach $\omega _{gq_{y}}.$ So these modes should be
considered as quasi-ripplonic oscillations which do not contribute to the
electron dynamics and, in particular, to the electron RMS displacement.

One should emphasize that the values of $g$ and $q_{y}$ that mainly
contribute to the results shown above satisfy the condition $g$ and $%
q_{y}\lesssim (\langle u_{f}^{2}\rangle )^{-1/2}$ due to the exponential
cutoff entering into the sums in Eqs. (\ref{9})-(11). We have also shown
that $L_{y}<R$.\cite{shs97} For $L_{y}=10^{-4}$ cm, ($q_{y})_{\min }=2\pi
/L_{y}>10^{4}$ cm$^{-1}$ and taking into account that ($g)_{\min }$ is of
the same order of magnitude, we obtain, for $(\langle u_{f}^{2}\rangle
)^{1/2}\lesssim 10^{-5}$ cm and for $a=10^{-4}$ cm, the values $\omega
_{gq_{y}}\sim 10^{7}-10^{8}$ s$^{-1}$ for the frequencies of ripplons
participating in the electron-ripplon coupling, which are significantly
smaller than the values of the threshold frequencies. Furthermore, the
low-frequency transverse mode $\stackrel{\thicksim }{\omega }_{t}$ given by
the first of Eqs. (\ref{11a}), reaches a value much lower than 10$^{8}$ s$%
^{-1}$. The maximum value of the longitudinal mode $\stackrel{\thicksim }{%
\omega }_{l}$ can be estimated by extrapolating Eq. (\ref{11a}) for $%
k_{x}=g_{1}=2\pi /a$ with the value of $\omega _{e}$ given by Eq. (\ref{1}).
The result is $3.06\times 10^{8}$ s$^{-1}$ and becomes smaller for $%
k_{x}\rightarrow 0$, where Eqs. (10) are valid. These results justify the a
priori assumption for taking the asymptotic limits $\omega ^{2}\ll \omega
_{gq_{y}}^{2}$ and $\omega ^{2}\gg \omega _{gq_{y}}^{2}$, in the long
wavelength limit, for the calculation of the dispersion laws of Eqs. (\ref
{11a}) and (\ref{13}). However one should note that using $\omega ^{2}\ll
\omega _{gq_{y}}^{2}$ for calculating mode spectrum $\stackrel{\thicksim }{%
\omega }_{l}$ may become questionable with increasing $k_{x}$ because in
this case the mode frequency can be of the same order of magnitude of $%
\omega _{gq_{y}}.$ In such a condition, $\stackrel{\thicksim }{\omega }_{l}$
should be calculated from a numerical solution of the first of Eqs. (\ref{10}%
). One can also point out that the splitting of low- and high-frequency
modes $\omega _{tr(l)}$ and $\omega _{tr(t)}$ is rather well-pronounced
supporting the introduction of a Debye-Waller factor to take into account
the smearing of electron positions from fast modes.

In conclusion we have calculated the coupled electron-phonon mode spectrum
of the single quasi-crystalline wire on the liquid helium surface. We have
shown interesting features of the spectrum, which makes it particularly
different from its 2D counterpart. Because the strong coupling of the
phonons of the WS wire with the oscillations of the helium surface, the
spectrum of collective modes changes drasticaly in comparison with the
decoupled wire electron system. The low-frequency (high-frequency) modes
have frequencies significantly smaller (larger) than characteristic
frequencies of ripplons which contribute to the electron-ripplon scattering.
Low-frequency modes have same dispersion as that of modes of the decoupled
electron wire but with large effective masses different for longitudinal and
transverse oscillations. The values of these effective masses are much
larger than the effective mass in the 2DWS that means that the
electron-ripplon coupling in Q1D case should be stronger. High frequency
longitudinal and transverse modes start from different threshold
frequencies. The value of the threshold frequency for the transverse mode is
slight affected by the electron-ripplon coupling and is close to that of
decoupled electron wire. In our opinion, experimental observation of the
predicted modes should give strong evidence of the phase transition of the
classical electron wire to a quasi-ordered state.

This work was partially sponsored by the Conselho Nacional de
Desenvolvimento Cientifico e Tecnol\'{o}gico (CNPq), Brazil.

FIG. 1. Structure of the single classical electron wire over liquid helium.
\end{mathletters}

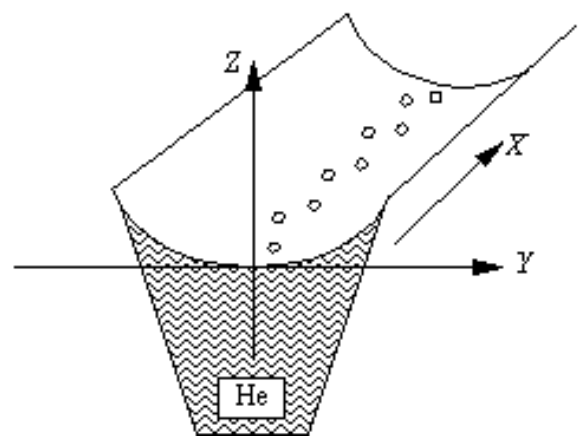
\end{document}